\documentclass[]{aa}   
\usepackage{graphicx}
\usepackage{multirow}
\usepackage{subcaption}

\begin{document}
\title{Phase mixing and wave heating in a complex coronal plasma}
\author{T. A. Howson \inst{1} \and I. De Moortel \inst{1, 2} \and J. Reid \inst{1}}
\institute{School of Mathematics and Statistics, University of St Andrews, St Andrews, Fife, KY16 9SS, U.K. \and Rosseland Centre for Solar Physics, University of Oslo, PO Box 1029  Blindern, NO-0315 Oslo, Norway}

\abstract{}
{We investigate the formation of small scales and the related dissipation of magnetohydronamic (MHD) wave energy through non-linear interactions of counter-propagating, phase-mixed Alfv\'enic waves in a complex magnetic field.}
{We conducted fully three-dimensional, non-ideal MHD simulations of transverse waves in complex magnetic field configurations. Continuous wave drivers were imposed on the foot points of magnetic field lines and the system was evolved for several Alfv\'en travel times. Phase-mixed waves were allowed to reflect off the upper boundary and the interactions between the resultant counter-streaming wave packets were analysed.}
{The complex nature of the background magnetic field encourages the development of phase mixing throughout the numerical domain, leading to a growth in alternating currents and vorticities. Counter-propagating phase-mixed MHD wave modes induce a cascade of energy to small scales and result in more efficient wave energy dissipation. This effect is enhanced in simulations with more complex background fields. High-frequency drivers excite localised field line resonances and produce efficient wave heating. However, this relies on the formation of large amplitude oscillations on resonant field lines. Drivers with smaller frequencies than the fundamental frequencies of field lines are not able to excite resonances and thus do not inject sufficient Poynting flux to power coronal heating. Even in the case of high-frequency oscillations, the rate of dissipation is likely too slow to balance coronal energy losses, even within the quiet Sun.}
{For the case of the generalised phase-mixing presented here, complex background field structures enhance the rate of wave energy dissipation. However, it remains difficult for realistic wave drivers to inject sufficient Poynting flux to heat the corona. Indeed, significant heating only occurs in cases which exhibit oscillation amplitudes that are much larger than those currently observed in the solar atmosphere.} 

\maketitle


\section{Introduction}\label{sec:introduction}
The dissipation of magnetohydrodynamic (MHD) wave energy has long been posited as a mechanism for maintaining the high temperatures observed within the solar corona \citep[for example, see reviews by][]{Erdelyi2007, Parnell2012, Arregui2015}. As a result of significant improvements in observational capabilities over recent years, many authors have been able to identify the prevalence of a vast array of MHD waves within the solar atmosphere \citep[for example, see reviews by][]{Ofman2005, Tripathi2009, Gallagher2011, Wang2011}. Detections of propagating transverse Alfv\'en and kink waves throughout the coronal volume are of particular interest to the current study \citep[e.g.][]{Tomczyk2007, Thurgood2014, Morton2015, Tiwari2019}. Whilst the energy budget associated with these oscillations is not well constrained, it remains possible that it is sufficient to balance expected energy losses, particularly in the quiet Sun \citep{DePontieu2007, McIntosh2011, Morton2012, Srivastava2017}. 

However, even with abundant oscillatory power, it remains unclear whether sufficient heating can occur on a time scale that is comparable to the rate of energy losses. Coronal plasma is associated with high Lundquist and Reynolds numbers and, as such, the rate of energy dissipation is typically expected to be very slow. Thus, in order to heat plasma on the time scales of radiative and conductive losses, wave energy must exist at small spatial scales within the solar atmosphere. To this end, many studies have highlighted mechanisms by which small scale variations may form within coronal wave modes. For example, large-scale, standing and propagating kink modes are able to efficiently transfer energy to small-scale, localised, Alfv\'en waves through the process of resonant absorption and mode coupling \citep{Ionson1978}. Subsequently, even smaller scales can form if the velocity shear that is associated with the localised Alfv\'en waves becomes unstable to the Kelvin-Helmholtz instability \citep[e.g.][]{Terradas2008, Antolin2014}. In this regime, wave energy cascades to the dissipation length scale before inevitably being converted into heat \citep[e.g.][]{Magyar2016, Karampelas2017}. Observational evidence for these effects are discussed in the pair of articles by \citet{Okamoto2015} and \citet{Antolin2015}. 

An alternative mechanism for generating turbulent-like flows (and hence small spatial scales) from coronal MHD waves is associated with the non-linear interaction of counter-propagating oscillations \citep[e.g.][]{VanBallegooijen2011, VanBallegooijen2017}. The different spatial profiles of the interacting wave packets allow the development of higher wavenumber modes and, ultimately, turbulent dissipation \citep{Shebalin1983, Oughton1994}. Several models have argued that the formation of such turbulence can lead to coronal heating and the acceleration of the fast solar wind \citep[e.g.][]{Oughton1999, Smith2001, Oughton2001, Dmitruk2002, Verdini2010, Woolsey2015, Shoda2019}. For incompressible plasmas, counter-propagating waves (possibly generated by reflections from a vertical (field-aligned) stratifcation of the Alfv\'en speed, for example) are required for this turbulence to develop. However, \citet{Magyar2017}, developed a model in which non-linear, unidirectional, MHD waves can trigger the formation of turbulence in plasmas with significant radial (cross-field) density structuring. In the aforementioned studies, wave energy experiences a turbulent cascade to the dissipation length scale, whereupon it increases the temperature of the plasma. 

Additionally, the existence of transverse gradients in the wave speed leads to the formation of large spatial gradients through the process of phase mixing. In a simple geometry with a straight, uniform magnetic field and a non-uniform density profile, \citet{Heyvaerts1983} showed that the heating rate is inversely proportional to the cube root of the magnetic Reynolds number, $R_M$. However, subsequent studies have highlighted that this does not provide an adequate enhancement in the heating rate to balance radiative losses unless artificially large dissipation coefficients are implemented \citep[e.g.][]{Cargill2016, Pagano2017, Prok2019b}. However, an analytic study by \citet{Similon1989} claimed that, within closed coronal arches, if the background magnetic field is sufficiently complex, the expected heating rate due to phase mixing is proportional to $\log R_M$. In terms of wave heating, this provides a significant improvement upon the classic case. Conversely, \citet{Parker1991} identified the lack of an ignorable co-ordinate in a fully 3-D filamentary coronal hole as an argument for weak wave heating in open-field environments. 

At the current time, the complexity of the magnetic field within the solar atmosphere is not well known. It has been hypothesised that the continuous convective driving of the photospheric surface ensures that the coronal field exists in a state of permanent stress. Indeed, an alternative view of coronal heating dictates that the impulsive release of this magnetic energy is responsible for maintaining coronal temperatures \citep{Parker1972}. In this regime, the coronal field must have a complex and intricate topology in which fine current sheets release energy through magnetic reconnection and Ohmic heating. 

If this does represent the true nature of the solar magnetic field, then the observed wave modes are likely to exist as perturbations to a complex background state and, as such, it is important to understand the dynamics and energetics of waves in this context. In \citet{Howson2019}, we established key characteristics of the propagation of a single wave pulse through a variety of magnetic field geometries with a range of complexities. In particular, we established that phase mixing is able to generate small spatial scales throughout the cross-section of the wave packet. Additionally, we noted that the polarisation of the wave is modified by the tangled nature of magnetic field lines and that the complexity of the background field is critical for determining the rate of energy dissipation.  

In the current article, we discern the effects of continuous wave driving and discuss the wave heating rate that may be expected in this regime. We present the results of 3-D numerical MHD simulations and investigate the effects of field complexity and viscous dissipation. In Section \ref{Num_Method}, we detail our model and discuss the initial conditions considered. In Section \ref{sec_Results}, we present our results and in Section \ref{sec_Discussion}, we discuss the implications of the findings of this study.

\section{Numerical method} \label{Num_Method}
For the simulations presented within this article, we advance the MHD equations using the Lagrangian-Remap code, Lare3d \citep{Arber2001}. The scheme advances the normalised MHD equations given by
\begin{equation}\frac{\text{D}\rho}{\text{D}t} = -\rho \vec{\nabla} \cdot \vec{v}, \end{equation}
\begin{equation} \label{eq:motion} \rho \frac{{\text{D}\vec{v}}}{{\text{D}t}} = \vec{j} \times \vec{B} - \vec{\nabla} P + \vec{F}_{\text{visc.}}, \end{equation}
\begin{equation} \label{eq:energy} \rho \frac{{\text{D}\epsilon}}{{\text{D}t}} = \eta \lvert \vec{j}\rvert^2- P(\vec{\nabla} \cdot \vec{v}) + Q_{\text{visc.}}, \end{equation}
\begin{equation}\label{eq:induction}\frac{\text{D}\vec{B}}{\text{D}t}=\left(\vec{B} \cdot \vec{\nabla}\right)\vec{v} - \left(\vec{\nabla} \cdot \vec{v} \right) \vec{B} - \vec{\nabla} \times \left(\eta \vec{\nabla} \times \vec{B}\right). \end{equation}
Here, $\rho$ is the mass density, $\vec{v}$ is the plasma velocity, $\vec{j}$ is the current density, $\vec{B}$ is the magnetic field, $P$ is the gas pressure, $\epsilon = P/\rho(\gamma - 1)$, is the specific internal energy (and $\gamma = 5/3$). In these equations, we have neglected the effects of thermal conduction, optically thin radiation and gravity.

The non-ideal terms, $\eta,$ and $\vec{F}_{\text{visc.}}$ and $Q_{\text{visc.}}$ are associated with the resistivity, acting on the magnetic field, and viscous contributions, acting on the velocity field, respectively. For the simulations presented hereafter, the resistivity, $\eta$ is set to 0. Consequently, the only user-imposed dissipation is in the form of viscosity, $\nu$. This exerts a force, $\vec{F}_{\text{visc.}}$, in the equation of motion (\ref{eq:motion}) and an associated heating term in the energy equation (\ref{eq:energy}). The viscous effects are a sum of contributions from a background viscosity, $\nu$, and two small shock viscosity terms which are included within all following simulations to ensure numerical stability. With the exception of the case $\nu = 0$, the effects of these terms are small in comparison to the background viscosity. Additionally, a small amount of numerical dissipation is inevitable in this finite difference scheme. Since Lare3d does not enforce energy conservation, this numerical dissipation removes energy from the domain and does not contribute to any heating term within the energy equation (\ref{eq:energy}).

\subsection{Initial conditions}
We seek initial conditions of coronal-like plasmas containing complex magnetic field geometries and a highly variable Alfv\'en speed profile. Additionally, in order to analyse the wave dynamics, we require the initial states to be close to (numerical) equilibrium. We follow the method presented in \citet{Howson2019} and allow simulation snapshots from the experiments presented in \citet{Reid2018} to relax to a numerical equilibrium under the effects of a large viscosity. 

\begin{figure}[h]
  \centering
  \includegraphics[width=0.5\textwidth]{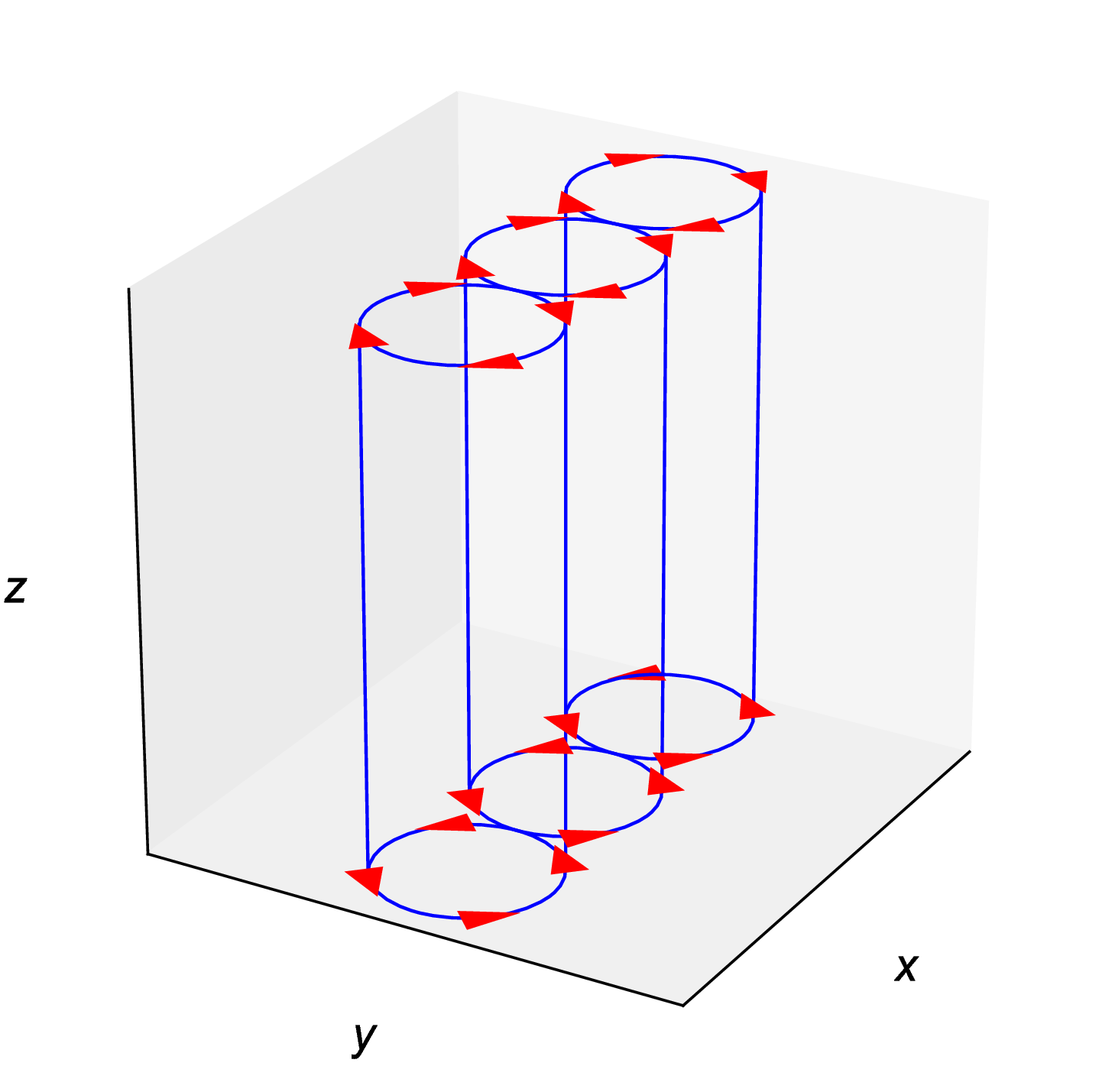}
  \caption{Schematic of the foot point driving implemented in \citet{Reid2018}. The red arrows indicate the direction of the velocity imposed at the foot points of the three magnetic threads.}
  \label{Drive_cartoon}
\end{figure}

We now provide a brief description of the state of the initial field. In \citet{Reid2018}, three counter-rotational, circular drivers are imposed on the upper and lower boundaries of a numerical domain that is initially threaded with a uniform magnetic field. The domain represents a closed field region of the corona, with magnetic foot points at both the upper and lower boundaries of the computational grid. These boundaries are located at the top of the transition region and are driven by motions propagating into the corona from lower layers of the atmosphere. In order to simplify the model, the curvature of field lines has been neglected. A schematic of the imposed driving is shown in Fig. \ref{Drive_cartoon}. 

The central thread is rotated at a faster rate than the other two and thus reaches the threshold for kink instability sooner. The development of this instability destabilises the remaining two threads and, under the action of continuous driving, generates a stressed magnetic field containing small scale current sheets which are widely distributed throughout the domain. In a non-ideal regime, this leads to Ohmic and viscous heating that converts the kinetic and magnetic energy injected by the rotational drivers into heat. 

Following the cessation of the foot point driving, the plasma remains in a stressed state and large forces generate significant flows throughout the domain. Therefore, in order to obtain suitable numerical equilibria, a large viscosity is introduced and the plasma velocities decrease until they are much smaller than the amplitude of the wave driver that is imposed in the current study (see below). Henceforth, we refer to this time immediately prior to the imposition of the wave driver as $t = 0$. During the numerical relaxation stage the resistivity, $\eta$ is set to 0 and thus (excluding numerical effects), we expect the magnetic connectivity to remain unchanged.  

We implement a numerical domain of dimensions 30 Mm $\times$ 30 Mm $\times$ 100 Mm and a grid consisting of 256 $\times$ 256 $\times$ 1024 cells. The initial equilibrium is sensitive to the spatial resolution of the finite difference scheme as gradients (e.g. currents for the magnetic field) form on smaller scales for finer numerical grids. As such, more intricate background currents are obtained if higher resolution simulations are conducted.

\begin{figure}[h]
  \centering
  \includegraphics[width=0.5\textwidth]{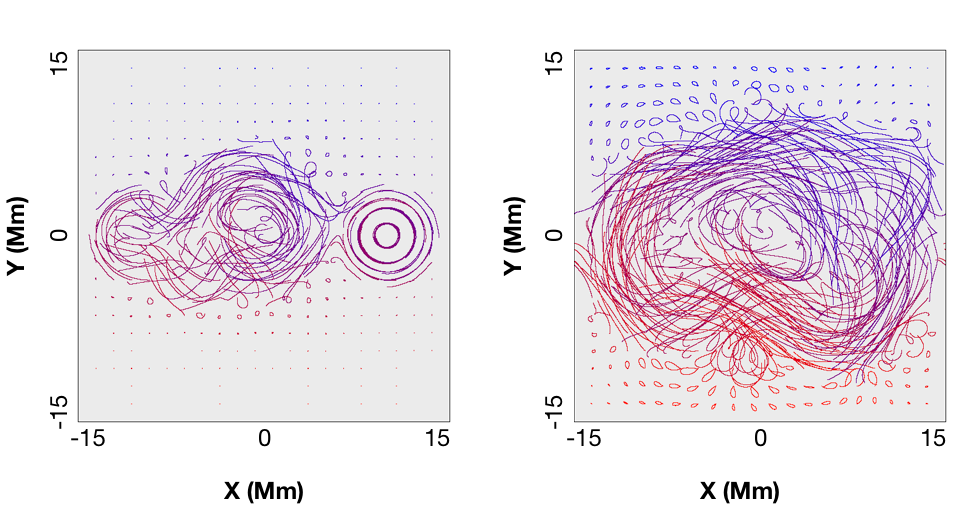}
  \caption{Projections of magnetic field lines onto $x$-$y$-planes. We show the two background magnetic fields investigated within this article. The less complex initial condition (s1) is on the left and the more complex one (s5) is on the right.}
  \label{In_Field_Top}
\end{figure}

\begin{figure}[h]
  \centering
  \includegraphics[width=0.5\textwidth]{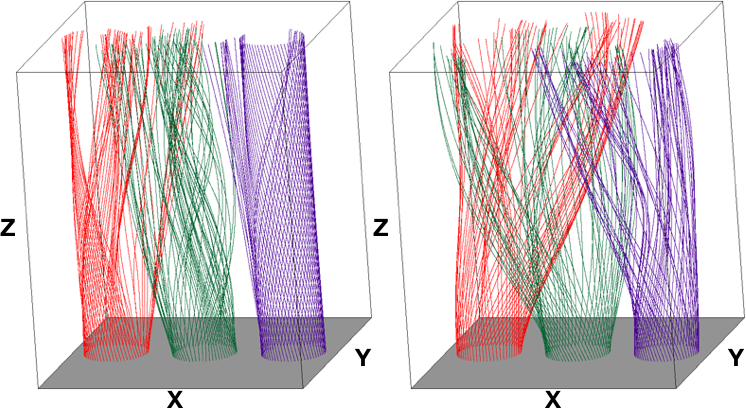}
  \caption{Magnetic field lines traced from the foot points of each magnetic thread. The less complex initial condition (s1) is on the left and the more complex one (s5) is on the right.}
  \label{In_Field_Side}
\end{figure}

Within this article, we consider two different initial conditions. These are obtained by allowing the rotational driving implemented in the \citet{Reid2018} study to proceed for different lengths of time. In one case, the driving continues for 100 Alfv\'en times and in the other, it continues for 500 Alfv\'en times. Hereafter, we refer to these initial conditions as the s1 and s5 cases, respectively. The numerically relaxed fields for each of these cases are shown in Figs. \ref{In_Field_Top} \& \ref{In_Field_Side}. In Fig. \ref{In_Field_Top}, we show the projection of magnetic field lines onto the lower $z$ boundary and in Fig. \ref{In_Field_Side}, we show field lines traced from three rings on the lower boundary, each corresponding to the location of the rotational driving (see Fig. \ref{Drive_cartoon}). In the case of the simpler field (left hand panel of Fig. \ref{In_Field_Side}), the central (green) and left hand (red) flux tubes have merged to form one magnetic structure, whereas the right hand (purple) flux tube remains distinct. In the case of the more complex field (right hand panel), on the other hand, all three flux tubes have been disrupted by the development of the kink instability. 

It is important to note that since the location of the rotational driving is confined to small values of $|y|$, in both cases, the final magnetic field is signficantly more complex close to $y=0$ (see Fig. \ref{In_Field_Top}). The heating studied by \citet{Reid2018}, and the further energy dissipation that occurs during the numerical relaxation, ensures the plasma parameters are different between the two cases. The state of the plasma is summarised in Table \ref{Tab_eqm_values}.

\begin{table*}[]
\begin{centering}
\begin{tabular}{c|ccc|ccc|ccc|ccc}
\multirow{2}{*}{Simulation} & \multicolumn{3}{c|}{Temperature (MK)} & \multicolumn{3}{c|}{Field Strength (G)} & \multicolumn{3}{c|}{Density ($\times 10^{-12}\text{ kg m}^{-3}$)} & \multicolumn{3}{c}{Plasma-$\beta$} \\ 
                  & Min.        & Mean       & Max.       & Min.        & Mean        & Max.        & Min.        & Mean        & Max.        & Min.       & Mean      & Max.      \\ \hline
 & & & & & & & & & & & & \\
s1               & 1.9         & 2.0        & 2.5        & 9.9         & 10.0        & 10.4        & 1.34         & 1.67         & 1.1         & 0.13       & 0.14      & 0.16      \\
s5              & 1.9         & 2.4        & 9.2       & 9.6         & 10.0        & 10.7        & 0.50        & 1.67         & 2.34        & 0.12       & 0.17     & 0.24      \\     
\end{tabular}
\caption{Values of plasma parameters for the initial conditions considered within this article. The quantities relate to the state of the plasma following the numerical relaxation and prior to the start of the wave dirivng.}
\label{Tab_eqm_values}
\end{centering}
\end{table*}

In each case, the post-relaxation state is highly inhomogeneous. In particular, there are small scales present in the density, magnetic field strength, and consequently, the local Alfv\'en speed. This, coupled with the non-constant length of magnetic field lines (due to the variable amount of magnetic twist within the domain), ensures that the natural Alfv\'en frequency of field lines varies significantly throughout the domain. The variability is typically larger within the s5 initial equilibrium as the field is more complex in this case.

\begin{figure*}[h]
  \centering
  \includegraphics[width=0.95\textwidth]{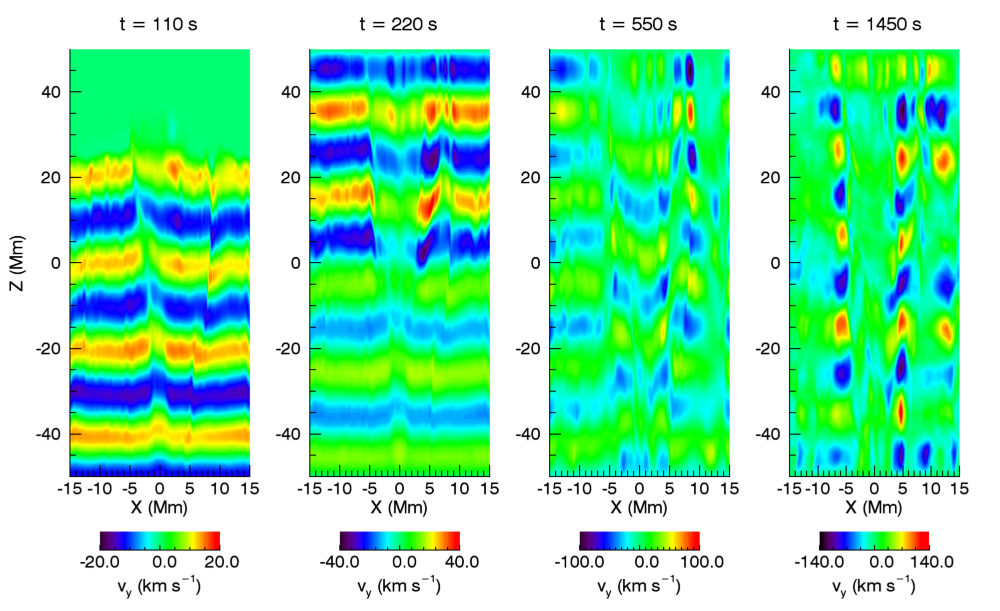}
  \caption{The $y$-component of the velocity in vertical slices through $y = 0$ Mm at four times during the simulation. The scale of the colour bar changes between each panel.}
  \label{vy_slices}
\end{figure*}

\subsection{Boundary conditions}
A continuous velocity driver, $\vec{v} = (0, v_y, 0)$ is imposed on the lower $z$ boundary, where, in the most basic case, $v_y$ is defined by 
\begin{equation} \label{eq:driver}
v_y = v_0 \sin{\omega t}.
\end{equation}
Here, the driver amplitude, $v_0 \approx 20 \text{ km s}^{-1}$ and, the frequency, $\omega = 0.21 \text{ s}^{-1}$, giving a time period of $\tau \approx 28$ s. We note that the driver amplitude is small in relation to the local Alfv\'en speed. This high frequency driver \citep[in comparison to the 3-5 minute periods often observed, e.g.][]{Tomczyk2009, Morton2016} is selected to ensure several wavelengths fit within the length of the magnetic structure (100 Mm). Additionally, these wave periods of a few tens of seconds are comparable to those observed in spicule oscillations \citep[e.g.][]{Okamoto2011, Yoshida2019}. Alternative forms of the imposed wave driving, including lower frequency oscillations, are discussed in detail below (see Sect. \ref{new_freq}).

Within this article, we assume that the background coronal plasma does not change over the duration of the simulations. In reality, low frequency photospheric motions modify the coronal field simultaneously with the high frequency disturbances being modelled here. We choose to separate these two mechanisms in order to isolate the effects of wave driving. This approach is not unreasonable given changes to the background field associated with long time scale driving are small over the course of a few wave periods, especially if high frequency waves are considered.

In all simulations the $x$ and $y$ boundaries are periodic \citep[as with the experiments presented in][]{Reid2018}. The upper boundary uses zero gradients for all variables, except for velocities which are set to 0. This ensures that the upper $z$ boundary acts as a mirror and wave packets that reach this boundary immediately reverse direction and return into the domain. The lower boundary also employs a zero gradient condition for all variables except for the velocity described above.
\section{Results}\label{sec_Results}

\begin{figure}[h]
  \centering
  \includegraphics[width=0.5\textwidth]{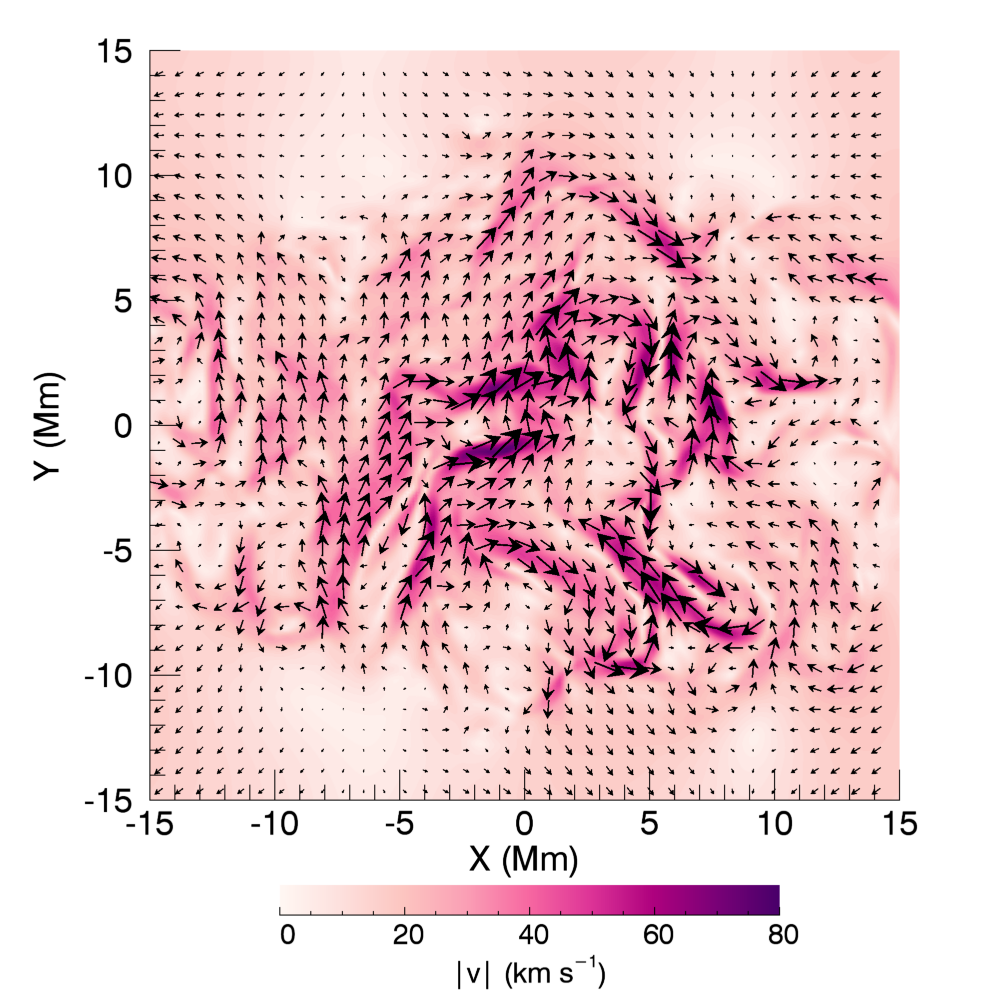}
  \caption{{\emph{Contour plot}} - Horizontal cut of the magnitude of the velocity in the midplane of the computational domain. \emph{Vector plot} - Horizontal velocity field. The time shown is $ t = 550$ s.}
  \label{vy_hor}
\end{figure}

We begin our analysis by considering a continuous, high-frequency wave driver imposed on the most complex background field (s5; right-hand panel of Fig. \ref{In_Field_Top}). The time period of the oscillation is 28 s and the mean Alfv\'en travel time of the domain is approximately 150 s. This ensures the domain contains approximately five wavelengths.

On account of the non-uniformity of the background plasma, the continuous driver excites a train of wave fronts that predominantly propagate along magnetic field lines and have a mixture of Alfv\'en and fast wave properties. The existence of transverse gradients in the local Alfv\'en speed and the variation in the length of field lines ensures that phase mixing distorts the wave packet as time progresses. Additionally, the existence of currents in the background magnetic field locally modify the polarisation of the wave front. Together, these processes generate small spatial scales in the velocity and perturbed magnetic fields which, in a non-ideal regime, enhance the rate of wave energy dissipation. These effects are discussed in detail in \citet{Howson2019} in the context of a single wave pulse and in this case, they simply occur for each wave front.

Once the leading wave packet reaches the upper boundary of the domain, it is allowed to reflect and the return waves interact non-linearly with the upwardly propagating waves. At this stage, it is important to note that the properties of a wave front at the reflecting boundary are very different to the spatially uniform mode that is driven from the lower $z$ boundary. As such, the non-linear interaction of these different counter-propagating waves are extremely complex, for example as discussed by \citet{VanBallegooijen2011}. We will show that this leads to a cascade of energy to smaller spatial scales and, in turn, for non-ideal plasmas, an enhanced heating rate.

In Fig. \ref{vy_slices}, we show the form of the wave fronts at four different times during the simulation. The first panel shows a time before the leading wave has reached the upper boundary. The effects of phase mixing are evident in the distortion of the wave fronts in the $x$ direction. In particular, it is clear that the leading wave front (which has experienced the most phase mixing) at approximately $z = 20$ Mm, is deformed to a much greater extent than the wave fronts close to the lower boundary. 

By the time of the second panel, the first wave fronts have reflected at the upper boundary and have started to interact with the counter-propagating waves that are continuously being excited by the imposed driver. In the remaining panels, a complex pattern of constructive and destructive interference is evident and small spatial scales form throughout the domain. In particular, as the simulation progresses, we see an increase in horizontal gradients in $v_y$ which contributes to the $z$-component of the vorticity (see below).

Since the background field is highly inhomogeneous in terms of the magnetic field strength, the density and the length of magnetic field lines, a wide range of natural Alfv\'en frequencies exists within the initial conditions. In particular, it is highly likely that for any given driving frequency, $\omega$ (larger than a typical fundamental frequency), we will find some field lines with a natural frequency (for some harmonic and polarisation) equal to $\omega$. Consequently, given that we have implemented a single-period, high-frequency, sinusoidal driver, we should expect to observe field line resonances. Indeed, in the third and fourth panels of Fig. \ref{vy_slices}, we see that very large velocities ($> 100 \text{ km s}^{-1}$) are excited over narrow regions, indicating the location of resonant field lines. For example, narrow resonant layers can clearly be observed at $x \approx -7$ Mm and $x \approx 5$ Mm.

The maximum observed velocities are almost an order of magnitude larger than the driver amplitude and certainly larger than any wave amplitudes that have been observed in the solar corona. Even though line-of-sight integration effects and limited spatial resolution may hinder observations of resonant coronal field lines (particularly for resonances confined to very narrow layers), if such large velocities did exist then they should be evident in non-thermal line broadening \citep[such as in][for example]{Brooks2016}. Alternatively, the amplitude of resonant oscillations could be limited by some unexpectedly high dissipation process and an associated heating of local plasma. Of course, it seems more likely that any waves excited by flows in the lower solar atmosphere are not driven by a single, high-frequency sinusoidal driver and this is explored in more detail in Sect. \ref{new_freq}.        

In these simulations, the magnitude of the velocities cannot increase indefinitely and is limited by physical (Ohmic and viscous) or numerical dissipation. In a non-ideal regime, the growth of the resonance and associated small scales in the perturbed velocity and magnetic fields will enhance the rate of wave energy dissipation (see Sect. \ref{non-ideal}). Additionally, non-linear effects, such as the redistribution of density along magnetic field lines by ponderomotive forces, will modify the natural Alfv\'en frequency of the field line and thus limit the size of any resonance \citep[see][for example]{Prok2019}.   

In Fig. \ref{vy_hor}, we show the behaviour of the velocity field in the horizontal midplane ($z=0$ Mm) at $t=550$ s. This time corresponds to the third panel of Fig. \ref{vy_slices}. The coloured contours show the magnitude of the velocity and the vectors (arrows) show the magnitude and direction of the horizontal velocity components, $v_x$ and $v_y$. The small spatial scales apparent in this figure are again indicative of the complex phase mixing. The largest flows are significantly larger than the amplitude of the driver and coincide with the location of resonances.

The imposed wave driver is incompressible, however, the velocity field at $z= 0$ Mm is certainly not. The compressibile nature of the velocity field is associated with a clear observational signature, which will be discussed in a forthcoming publication. Furthermore, the compression of the plasma also encourages the formation of non-linear fast-like MHD modes. These propagate across the field and transfer wave energy from resonant field lines to the neighbouring plasma. This can also act to reduce cross-field velocity gradients and, ultimately, reduce the energy dissipation efficiency of phase mixing.

In Fig. \ref{vy_hor}, we highlight the difference in the complexity of the velocity field between $|y| > 10$ Mm and $|y| < 10$ Mm. For small values of $|y|$, the background field is more complex (see Fig. \ref{In_Field_Top}) and thus the wave fronts become significantly more distorted than for large values of $|y|$. Since the wave modes are much more compressible close to the centre of the domain than close to the $y$ boundaries, synthetic observations are able to detect the regions of more complex field. This suggests the possibility of observations of propagating waves being used to deduce the complexity of the background coronal field.

\begin{figure*}[h]
  \centering
  \includegraphics[width=0.95\textwidth]{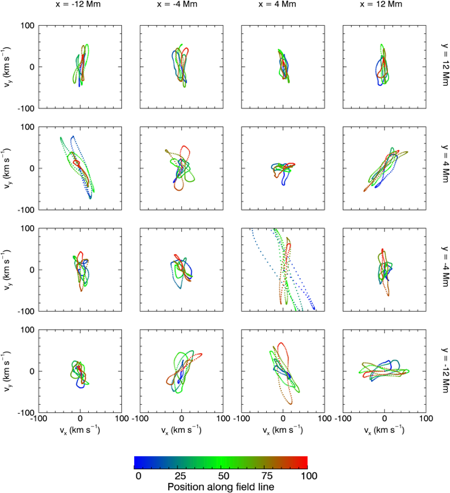}
  \caption{Polarisation of oscillations for different field lines within the domain at $ t = 1100$ s. Sixteen field lines are selected on a grid at positions indicated by the labels at the top and right-hand side of the figure. Three hundred points are selected along each field line and depicted with a colour that shows their position along the field line as a percentage of the total field line length. In order to show the direction of polarisation, the horizontal velocity ($v_y$ against $v_x$) is shown for each point. Notice the large resonance for row 3 column 3.}
  \label{polarisation_panel}
\end{figure*}

It is clear from Fig. \ref{vy_hor} that the excited oscillations do not all share the same polarisation as the imposed wave driver. Instead, the approximately helical nature of the initial magnetic field, and the associated background currents, induce a modification in the polarisation angle. This is a highly localised effect and generates a spectrum of polarisations throughout the numerical domain. 

In Fig. \ref{polarisation_panel}, we display a selection of these polarisations. The location of each panel within the grid corresponds to the location of the foot point (on the lower $z$ boundary) of a magnetic field line within the simulation. Within each panel, we show the $x$ and $y$ components of the velocity for 300 points along this magnetic field line. The colour of the points depicts the position of each point along a field line. Blue points are close to the lower $z$ boundary, green points are close to the midpoint of the field line and red points are close to the upper $z$ boundary. The figure shows the time $t = 1100$ s and by this point, large resonances have been generated within the simulation. 

Panels with points lying on or close to a straight line in velocity space (such as row 1 column 1) show field lines oscillating with a single linear polarisation. On the other hand, the second and third panels in the second column show field lines oscillating with a helical polarisation. This is simply caused by wave fronts interacting with the twisted magnetic structures within the complex field region. Even in regions with little background field complexity, fast-like waves emanating from the complex region are able to disturb the wave driven by the boundary motions, possibly resulting in significant variations in the oscillation angle.

Fig. \ref{polarisation_panel} also highlights the existence of localised resonances. For example, the third panel in the third column shows significantly larger velocities than are generated in much of the rest of the domain. Even on this resonant (or nearly resonant) field line, we see that the amplitudes of the antinode velocities change along the length of the field line. In particular, the antinodes with larger amplitudes (green and blue points) are located in regions of lower density and, consequently, higher Alfv\'en speed.

\begin{figure}[h]
  \centering
  \includegraphics[width=0.5\textwidth]{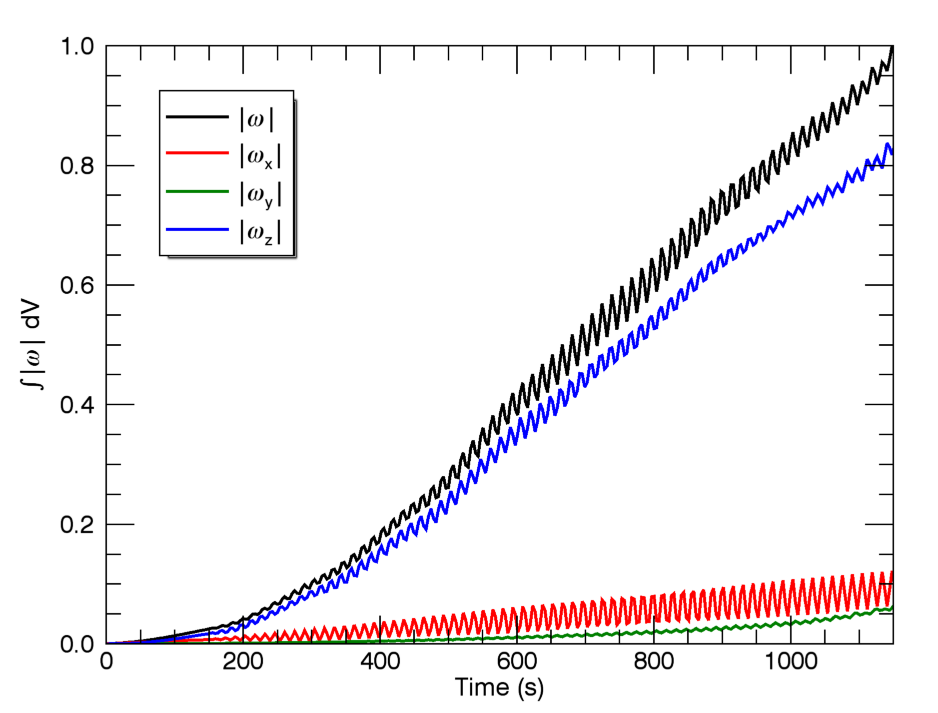}
  \caption{Volume integrated vorticity and the contribution of each component.}
  \label{vort_growth_og}
\end{figure}

The complex combination of phase mixing, the formation of localised resonances and the widespread modification of the polarisation angle enhances the formation rate of small scales within the simulation. When integrated over one time period, the wave energy is partitioned into kinetic and magnetic energy. The cascade of this wave energy onto small scales enhances viscous and Ohmic heating in the presence of non-zero viscosity and resistivity, respectively. 

As in \citet{Howson2019}, we utilise the volume integral of the magnitude of the vorticity, $|\omega|$ as a measure of the small scales present within the domain. An alternative metric may consider the volume integrated current density. However, unlike the vorticity, this is initially large due to the existence of background currents, and slowly diffuses due to numerical effects. Thus it would be more difficult to isolate the effects of the phase mixing waves. 

In Fig. \ref{vort_growth_og}, we show the time evolution of the volume integral of the magnitude of the vorticity during the simulation. The black line corresponds to the total vorticity, and the red, green and blue lines show the magnitude of the $x$, $y$ and $z$ components, respectively. The small amplitude oscillations that are observed in each line correspond to the frequency of the imposed wave driver.

The $x$ component of the vorticity is associated with vertical gradients in $v_y$ and is therefore related to the wavelength of the driven wave. The volume integral gradually increases as field line resonances enhance the amplitude of the waves and thus the $\frac{\partial{v_y}}{\partial{z}}$ contribution. Since the modification of the wave polarisation induces the growth of $|v_x|$ within the domain, the $y$ component of the vorticity is increased by the growth of the $\frac{\partial{v_x}}{\partial{z}}$ term. Additionally, the localised, non-linear generation of slow modes by ponderomotive forces has an effect on these two components of the vorticity. 

The main contribution to the total vorticity is the $\omega_z$ term (blue curve in Fig. \ref{vort_growth_og}). Despite the complexity of the background field, we have $|B_x| \sim |B_y| \ll |B_z|$. Therefore, the $z$ component of the vorticity is dominated by contributions from horizontal gradients in the transverse components of velocity. These gradients and, consequently, $|\omega_z|$ increase significantly as the phase mixing progresses and localised resonances form.

\begin{figure}[h]
  \centering
  \includegraphics[width=0.5\textwidth]{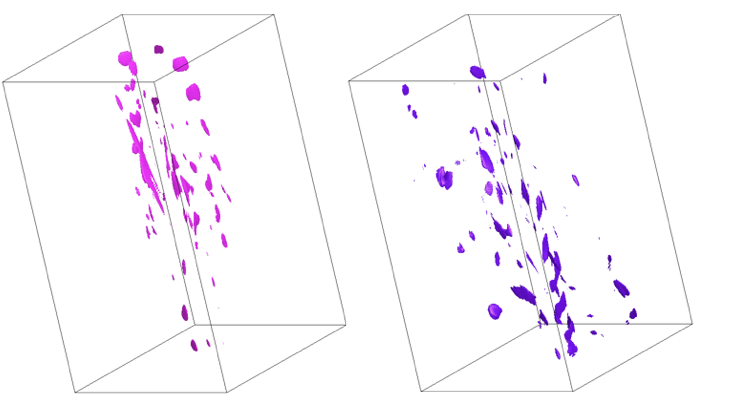}
  \caption{Isosurfaces of the instantaneous magnitude of vorticity (left) and current density (right) at $t = 1000$ s. The surfaces contain regions of high vorticity and current density, respectively.}
  \label{3d_cur_vort}
\end{figure}

Whilst the volume integrated vorticity increases gradually during the simulation, the spatial distribution of large velocity gradients changes rapidly over the oscillation period. The same can be said of the current density, although this is typically out-of-phase with the vorticity as the kinetic and magnetic wave energy components are also out-of-phase.

In Fig. \ref{3d_cur_vort}, we display isosurfaces of the magnitude of the vorticity (left-hand panel) and of the current density (right-hand panel) at a time of $t = 1000$ s. We note that by this time, the largest currents within the computational domain are associated with the wave dynamics and not the background magnetic field. The largest gradients form within the complex field region, where the effects of phase mixing are largest.

 Whilst the smallest scales do not form everywhere, in a non-ideal regime, plasma is heated relatively efficiently over a large volume of the domain. The structures observed in Fig. \ref{3d_cur_vort} are short-lived but form frequently and if they are associated with energy dissipation, are reminiscent of nanoflare heating events. However, the energy that is available to be dissipated within these events is a couple of orders of magnitude smaller than that assumed in typical nanoflare heating models \citep[e.g.][]{Reale2014, Klimchuk2015}. We note that despite the bursty, intermittent nature of the current and vorticity formation, any plasma heating might be interpreted as a steady process.

\subsection{Modifying the field complexity}
Hitherto, we have noted that the formation of small scales is much reduced in regions of the numerical domain which contain less complex field and, consequently, a more homogeneous Alfv\'en speed profile. Here, we examine the effects of the complexity of the initial (numerical) equilibrium. We present the results of the s1 simulation in which the same, high-frequency wave driver is imposed on the field displayed in the left-hand panel of Fig. \ref{In_Field_Top}. 

\begin{figure}[h]
  \centering
  \includegraphics[width=0.5\textwidth]{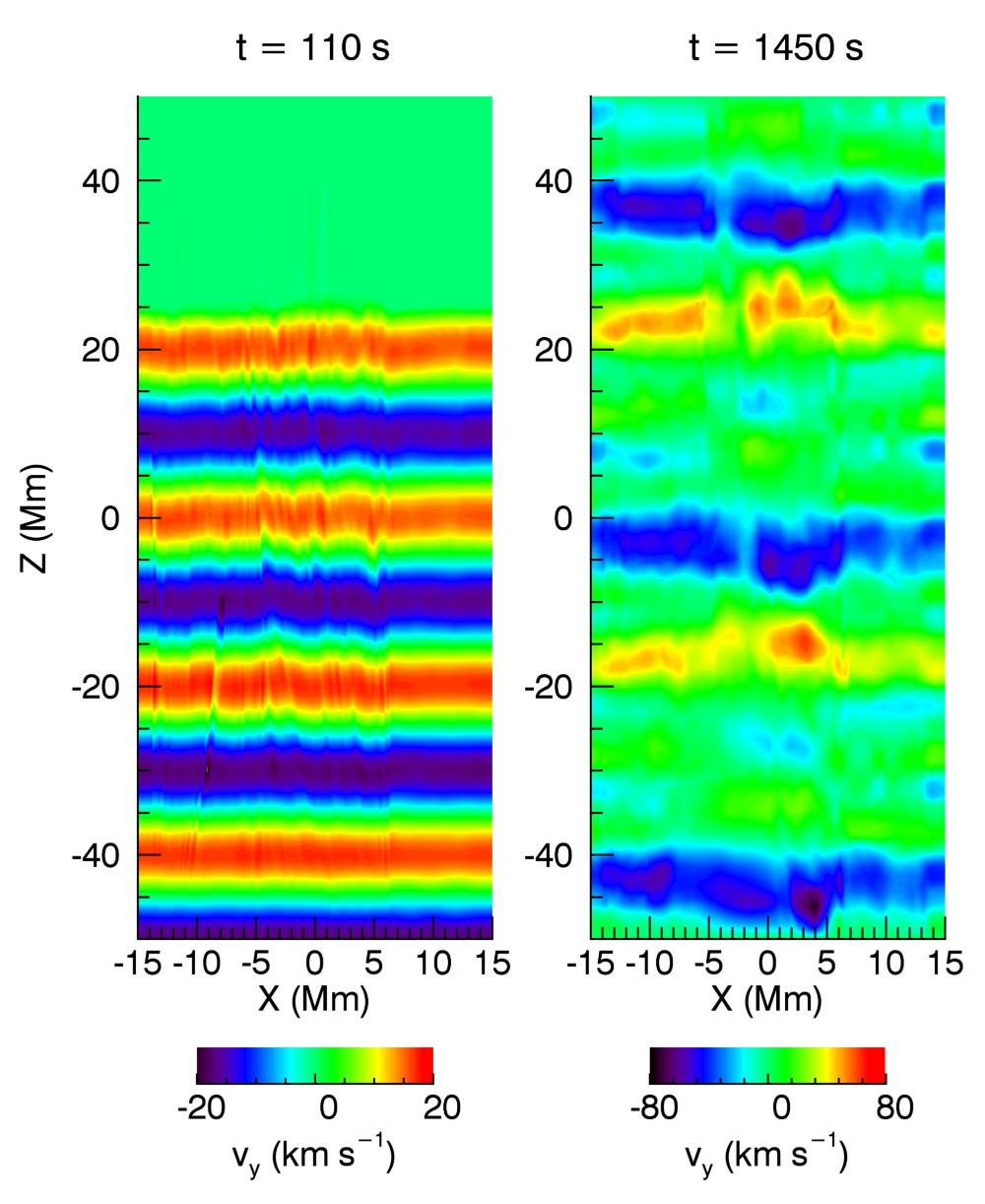}
  \caption{The $y$-component of the velocity in vertical slices through $y = 0$ Mm at two times during the simulation with the less complex initial configuration (s1). The scale of the colour bar changes between the two panels.}
  \label{20vslices}
\end{figure}

In Fig. \ref{20vslices}, we show the $y$ component of the velocity at two times, $t = 110$ s (left) and $t = 1450$ s (right) in the plane $y = 0$ Mm. The times shown correspond to the left- and right-most panels in Fig. \ref{vy_slices}. In both panels, we notice that the wave front remains much more coherent in the less complex field case. This is indicative of reduced phase mixing in a more uniform background plasma and magnetic field.

Since the natural Alfv\'en frequency of field lines varies less in this simulation, unless the driving frequency is close to the average of the natural field line frequencies, fewer resonances are excited. Despite this, large velocities (in comparison to the imposed driver) of $\sim 80 \text{ km s}^{-1}$ are still excited within this simulation. These form when couter-propagating waves experience constructive interference. They are much less localised than in the s5 simulation as neighbouring field lines are typically more similar in this case. 

\begin{figure}[h]
  \centering
  \includegraphics[width=0.5\textwidth]{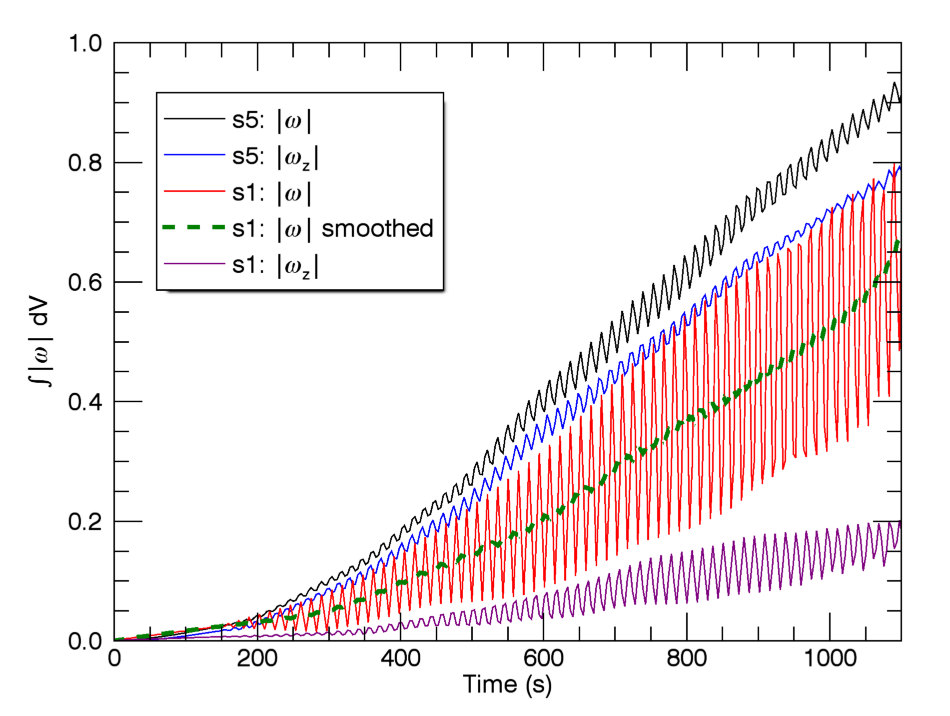}
  \caption{Volume integrated vorticity for the different initial fields (black and red). We also include a time-running average for the less complex initial equilibrium (green). The blue and purple curves show the vertical component of the vorticity and are a measure of the phase mixing for the two field configurations.}
  \label{20v100_vort}
\end{figure}

In Fig. \ref{20v100_vort}, we show the volume integral of the magnitude of $\omega$ and $\omega_z$ for the s1 (red and purple lines, respectively) and s5 (black and blue lines, respectively) simulations. The green curve shows a time-running average of the total vorticity for the s1 simulation (red curve). The black and blue lines are identical to those shown in Fig. \ref{vort_growth_og} and are included for comparison. 

It is clear that the vorticity, and hence small scales in the velocity field, are reduced when the wave driver is imposed on the less complex field. This is largely a result of the reduction in phase mixing which can be observed by comparing the vertical component of the vorticity between the two simulations (blue and purple curves). In particular, the horizontal gradients in the transverse velocity components are significantly smaller in the experiment with the simpler field (also compare Figs. \ref{vy_slices} \& \ref{20vslices}). There is an additional effect which is a consequence of a change in the resonances that form. Since the background plasma is more homogeneous in the s1 case, unless a resonant driving frequency is implemented, fewer resonant field lines form and thus the observed wave amplitudes and associated velocity gradients are reduced.     

In the original case, at any given time some field lines experience constructive wave interference whilst others show destructive wave interference. However, in this case, the increased uniformity of the background plasma ensures large volumes of the plasma experience constructive and/or destructive interference at the same time. This creates the large amplitude oscillation in the volume integrated vorticity (compare the red and black lines in Fig. \ref{20v100_vort}).

\subsection{Modified driving frequency} \label{new_freq}
\begin{figure}[h]
  \centering
  \includegraphics[width=0.5\textwidth]{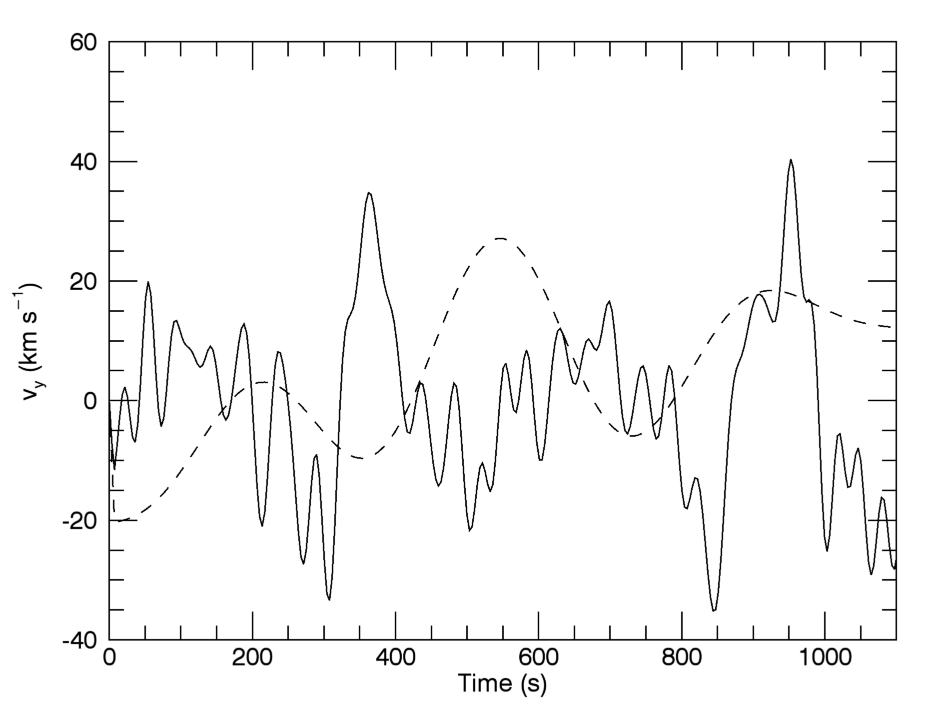}
  \caption{Velocity of the imposed wave driver for the r5a (solid line) and r5b (dashed line) simulations.}
  \label{rd_driver_freq}
\end{figure}

The wave driver imposed in the above simulations has a high, constant frequency and is unlikely to be representative of quasi-periodic perturbations in the lower solar atmosphere. In this section we consider the effects of modifying the driver such that it is no longer described by a simple sine wave. On the lower $z$ boundary, we impose a velocity of the form $\vec{v} = \left(0, v_y, 0\right)$ where $v_y = v_y(t)$ is defined using a sum of $N=1000$ sine waves with frequencies, $\omega_n$ between a maximum frequency $\omega_M$ and a minimum frequency, $\omega_M/10$. We consider two frequency ranges by setting $\omega_M = 0.21 \text{ s}^{-1}$, (the driving frequency implemented in previous sections) and $\omega_M =0.021 \text{ s}^{-1}$. We refer to these simulations as $r5a$ and $r5b$, respectively. In each case, every sine in the summation is given a phase, $\phi_n$, that is randomly selected from a uniform distribution on the interval $[0, 2\pi]$. Additionally, each term is associated with an amplitude that is proportional to $\omega_n^{-5/6}$. This is designed to generate a velocity driver with energy following a Kolmogorov power spectrum over this narrow frequency range. We have,    
\begin{equation} \label{new_driver}
v_y = v \sum_{n=1}^{N} \omega_n ^{-5/6} \sin\left(\omega_n t + \phi_n\right),
\end{equation}
where
\begin{equation}
\omega_n = \frac{\omega_M}{10}\left(1 + \frac{9n}{N}\right),
\end{equation}
and $v$ is a constant selected such that the mean of $|v_y|$ is approximately equal to the same quantity for the high-frequency driver described previously. Consequently, the smallest frequencies and longest time periods are associated with the largest energies. The forms of the drivers for the r5a and r5b simulations are displayed in Fig. \ref{rd_driver_freq}. The solid line corresponds to the higher frequency simulation (r5a) and the dashed line corresponds to the lower frequency simulation (r5b). It is important to note that the r5a simulation has wave power at frequencies above that of the fundamental modes of field lines within the domain, whereas the r5b simulation does not. 

\begin{figure}[h]
  \centering
  \includegraphics[width=0.5\textwidth]{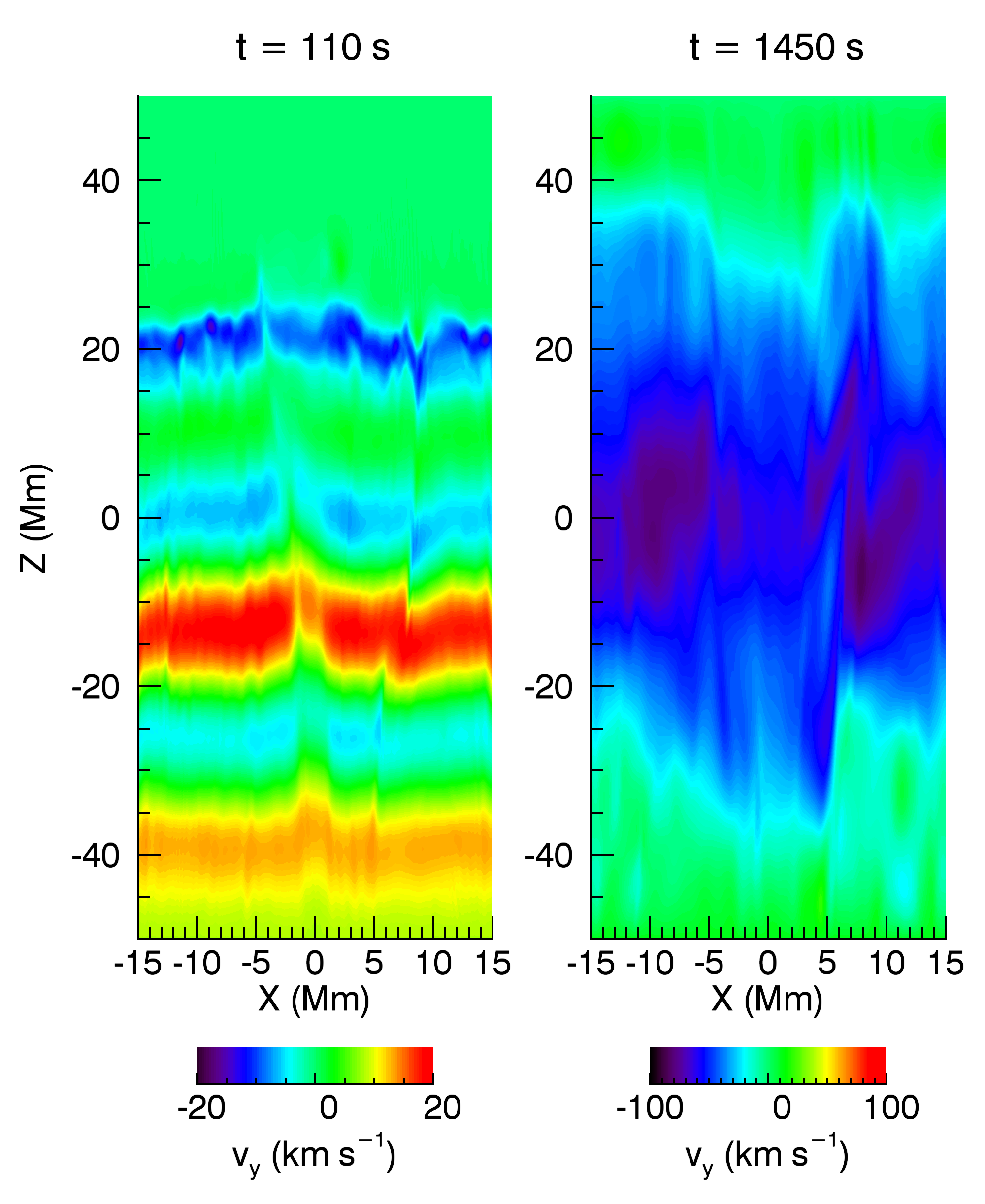}
  \caption{The $y$-component of the velocity in vertical slices through $y = 0$ Mm at two times during the r5a simulation. The scale of the colour bar changes between the two panels.}
  \label{rd_vslices}
\end{figure}

\begin{figure}[h]
  \centering
  \includegraphics[width=0.5\textwidth]{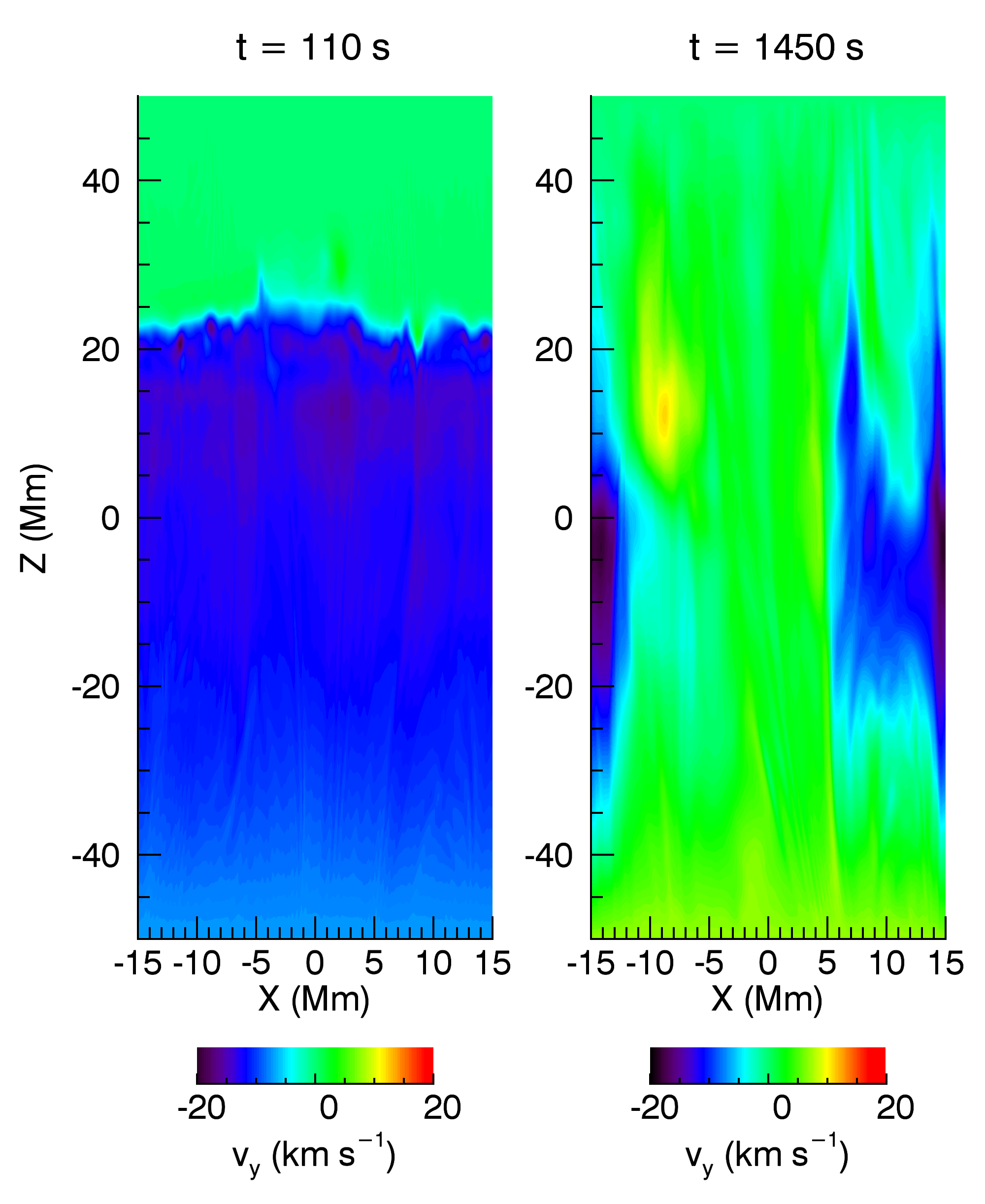}
  \caption{The $y$-component of the velocity in vertical slices through $y = 0$ Mm at two times during the r5b simulation.}
  \label{rd_vslices_lf}
\end{figure}

In Figs. \ref{rd_vslices} and \ref{rd_vslices_lf}, we show the $y$ component of the velocity induced by the modified driver for the r5a and r5b simulations, respectively. In each figure, we show two times, $t = 110$ s (left) and $t = 1450$ s (right) in the plane $y = 0$ Mm. The times shown correspond to the left- and right-most panels of Fig. \ref{vy_slices} and are the same as those shown in Fig. \ref{20vslices}. Whilst the new drivers are defined using the sum of sine waves, each has a lower frequency than in the original case and thus the vertical wavelengths are increased in these simulations. This is especially true in the r5b simulation, where the vertical wavelengths associated with the constituent sine waves are longer than the height of the computational box. In the right-hand panels of Figs. \ref{rd_vslices} and \ref{rd_vslices_lf}, we notice that the low frequency perturbations begin to dominate at later times. This is due to the greater proportion of wave power that is present at the lower frequencies (see equation \ref{new_driver}). 

\begin{figure}[h]
  \centering
  \includegraphics[width=0.5\textwidth]{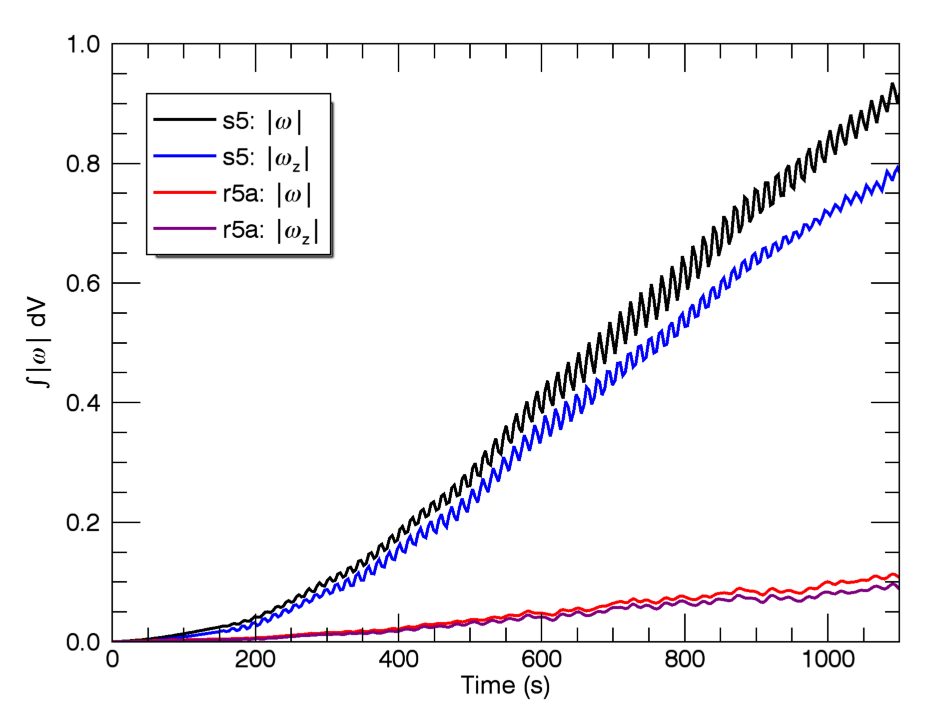}
  \caption{Volume integrated vorticity for the s5 (black curve) and r5a (red curve) simulations. The blue and purple curves show the vertical component of the vorticity and are a measure of the phase mixing for the two wave drivers.}
  \label{rd_vort_plot}
\end{figure}

In the right-hand panels of Figs. \ref{rd_vslices} and \ref{rd_vslices_lf}, again we notice the formation of small scales within the velocity field. However, as a result of the increased wavelengths, there must be a greater variation in the local Alfv\'en speed and/or field line length in order for significant phase mixing to occur. Therefore, whilst small scales still form in these simulations (particularly in the complex field region), the cascade of energy to the dissipation length scale is much slower. To this end, in Fig. \ref{rd_vort_plot}, we show the volume integral of the magnitude of the vorticity for the r5a simulation (red curve) and, for comparison, we also show the same quantity for the high-frequency, s5 simulation (black curve). We also include the integral of the magnitude of $\omega_z$, the component most sensitive to phase mixing. Clearly, the total vorticity is significantly reduced in the low frequency simulation and the reduced small scale formation rate can be attributed to the reduced phase mixing that occurs. For clarity, we have not included the corresponding curves for the r5b simulation, however, we note that the volume integrated vorticity is smaller than in the r5a case.

\subsection{Poynting flux}

\begin{figure}[h]
  \centering
  \includegraphics[width=0.45\textwidth]{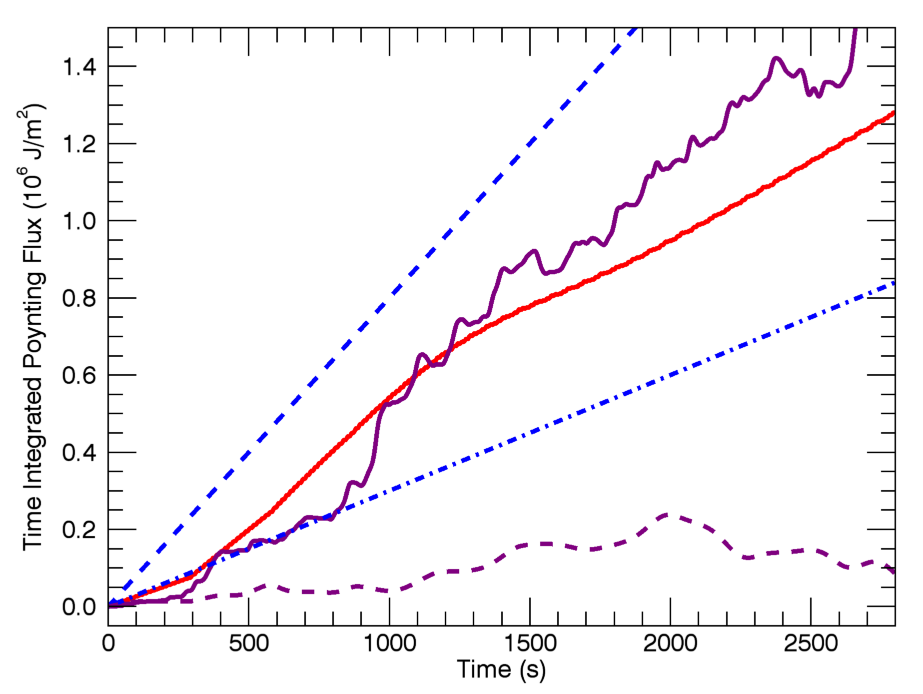}
  \caption{Time-integrated Poynting flux injected by the wave driver in the s5 (red), r5a (solid purple) and r5b (dashed purple) simulations. The blue lines indicate estimates of the energy requirements in the quiet sun (dot-dashed line) and for a coronal hole (dashed line).}
  \label{Poynt_flux}
\end{figure}
 
Of course, the total vorticity can also be influenced by the amount of wave energy injected into the domain. Thus, in this section, we compare the Poynting flux associated with the drivers described above. The existence of resonant field lines within the domain enhances the energy flux through the simulation boundary. Consequently, the amount of energy available for wave heating is sensitive to the frequency (as well as the amplitude) of the driver \citep[e.g.][]{Prok2019}. 

In Fig. \ref{Poynt_flux}, we display the spatially-averaged, time-integrated Poynting flux through the lower $z$ boundary over the course of the s5 (red line), r5a (solid purple line) and r5b (dashed purple line) simulations. The instantaneous Poynting flux is given by the derivatives of each of these curves. We also include estimates of the energy requirements for the Quiet Sun (dot-dashed blue line) and for a coronal hole (dashed blue line). These are calculated using values predicted in \citet{Withbroe1977} which account for expected energy losses due to thermal conduction, optically thin radiation and the solar wind. We note that these assume typical coronal conditions and are not specifically calculated for the plasma parameters used within these simulations. We observe that the s5 and r5a drivers would potentially be able to heat the Quiet Sun, but do not supply sufficient energy to balance losses in a coronal hole or within active regions (curve not shown but losses are significantly larger than for a coronal hole). The r5b driver, on the other hand, does not even provide sufficient energy to heat the Quiet Sun.

For the s5 case, although the driver remains constant throughout the simulation, the Poynting flux does not. In particular, it is smallest for $t \lesssim 300 s$. During this stage of the simulation, no reflected waves have had the time to return to the lower $z$ boundary. Beyond this time, the driver interacts with the perturbed magnetic field and thus the energy injected is modified. For resonant field lines, the Poynting flux increases, whereas for other field lines, the driver is able to remove wave energy from the domain. At this stage, the time-integrated (over a wave period) Poynting flux becomes sensitive to the driving frequency. If the driver frequency matches the natural frequencies of a greater number of field lines, then the Poynting flux will increase. We see that for this simulation, the reflected waves act to increase the Poynting flux through the driven boundary (change in gradient at $t \approx 300$ s). However, there is also a decrease in the instantaneous (derivative of purple curve in Fig. \ref{Poynt_flux}) Poynting flux at around $t = 1200$ s. This change in behaviour is associated with the detuning of resonant field lines as a result of non-linear effects redistributing plasma and modifying the natural Alfv\'en frequencies to be different from the driver frequency \citep[see, for example][]{Prok2019}.  

For the r5a and r5b simulations, the driver is a sum of many contributing waves, each with a different frequency. In the case of the r5a simulation, the highest frequencies are larger than the fundamental modes of field lines. As such, resonances are still able to form throughout the domain. For the case of r5b simulation, however, the shortest time period of the constituent waves is longer than the Alfv\'en travel time, and thus wave resonances are not excited. Consequently, there is a significant difference in the Poynting flux of energy into the domain (compare the purple curves) even though the amplitude of the driving velocity is comparable (see Fig. \ref{rd_driver_freq}). 

We note that the difference in time scales between the r5a and r5b is typically used to characterise coronal heating models as {\emph{AC mechanisms}} (the time period of the wave driver, $\tau_d$, is less than the Alfv\'en travel time of field lines, $\tau_A$) or {\emph{DC mechanisms}} $\left(\tau_d > \tau_a\right)$. The much reduced Poynting flux for the r5b simulation (long time scales), however, does not provide evidence against DC heating models. In particular, an imposed velocity driver in a typical DC heating model slowly stresses the field, which is not the case for this simple driver.  

\begin{figure}[h]
  \centering
  \includegraphics[width=0.45\textwidth]{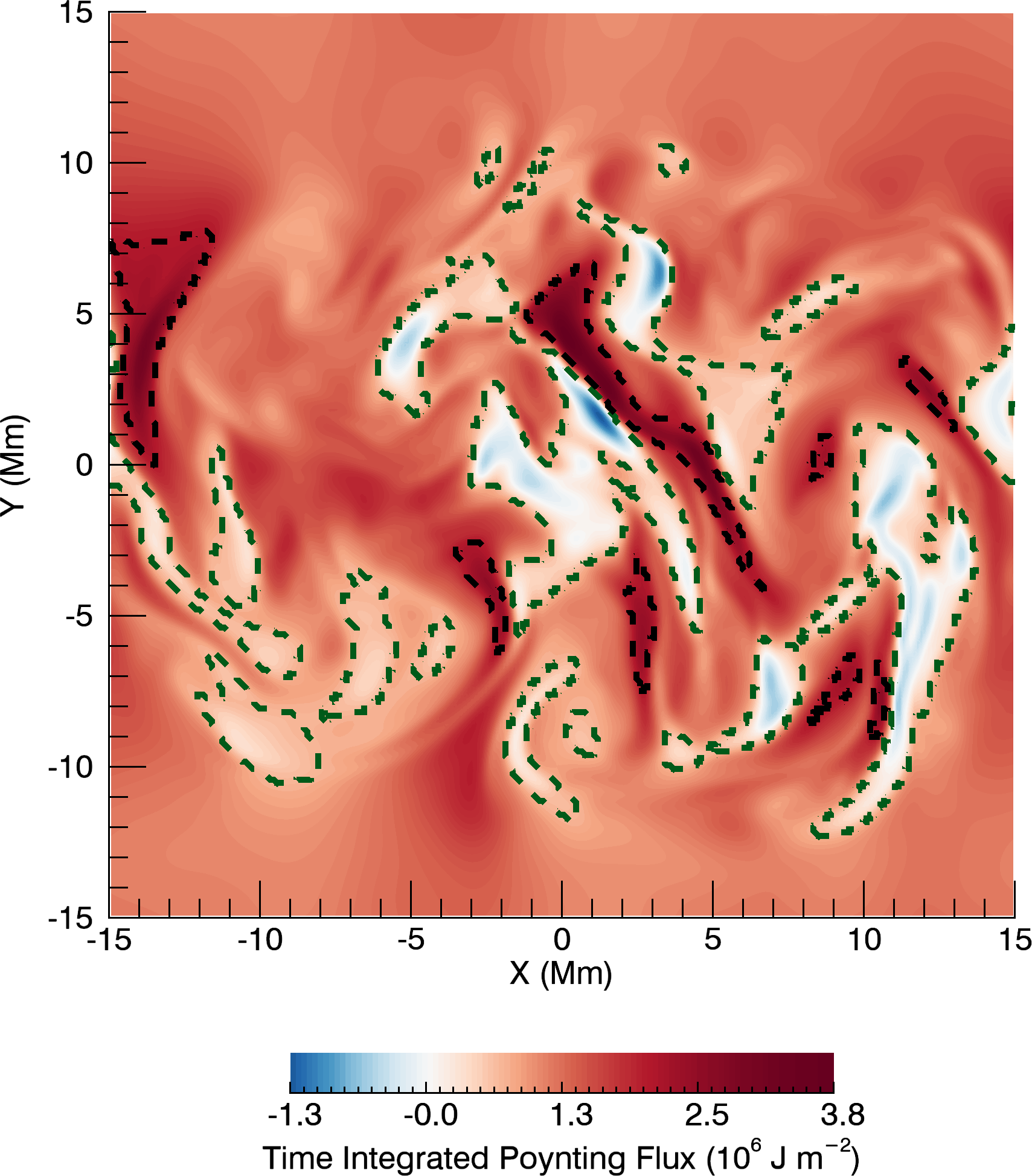}
  \caption{Time-integrated Poynting flux through the lower $z$ boundary over the course of the s5 simulation. Inside the black dashed contours, the Poynting flux is sufficient to power heating within a coronal hole. Inside the dark green dashed contours, the Poynting flux is not sufficient to heat the Quiet Sun.}
  \label{poynt_contour}
\end{figure}

In addition to changing throughout the duration of the simulation, the injected Poynting flux is also a function of space. In Fig. \ref{poynt_contour}, we display the time-integrated (over the duration the simulations) Poynting flux for the s5 simulation. Within the dashed dark green contour, the injected Poynting flux is not sufficient to heat the Quiet Sun \citep[according to][]{Withbroe1977}. The dashed black contour, on the other hand, contains all points through which the injected Poynting flux would be sufficient to balance energy losses in coronal hole conditions. However, we note that, since wave energy is not confined to field lines (on account of the formation of transversely propagating fast-like waves), this does not imply that field lines within the dark green contour could not be heated to a sufficient extent. For the r5a and r5b simulations, on the other hand, no natural Alfv\'en frequency is being preferentially selected by the broadband wave drivers. Consequently, the time-integrated Poynting flux through the lower $z$ boundary is much more uniform.

\subsection{Wave energy dissipation} \label{non-ideal}
Thus far, we have limited our consideration to ideal simulations and have not discussed the implications of the wave dynamics for plasma heating. In this section, we present the results of numerical experiments in which a non-zero viscosity is included in order to dissipate wave energy. We do not consider the effects of resistivity as this leads to the diffusion of the background field and results in the effects of the waves becoming difficult to isolate.

\begin{figure}[h]
  \centering
  \includegraphics[width=0.45\textwidth]{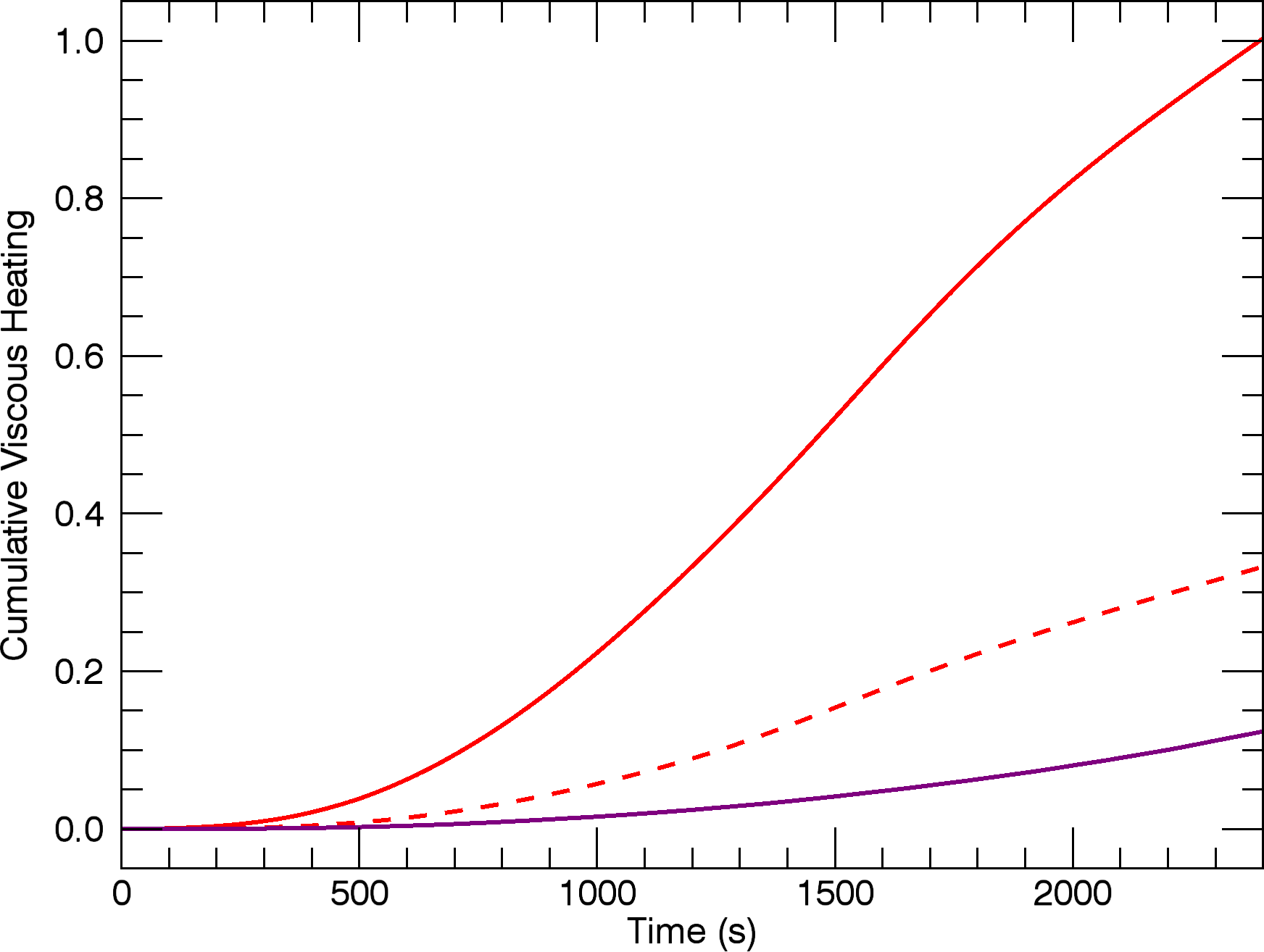}
  \caption{Cumulative viscous heating for the s5 simulation with Reynolds numbers of $\sim 10^3$ (solid red line) and $\sim 10^4$ (dashed red line) and for the r5a simulation with a Reynolds number of $\sim 10^3$ purple line. Here we have normalised by the total heating in the solid red curve simulation.}
  \label{visc_heating}
\end{figure}

We consider simulations with Reynolds numbers in the interval $[20, 10^5]$. We note that all values within this interval likely represent dissipative effects that are significantly larger than those present within the corona. However, with the exception of the largest values, they ensure that energy is dissipated before it cascades to scales at which the numerical dissipation becomes relevant. These simulations provide upper bounds on the plasma heating that we can expect in this regime. We restrict our analysis to waves within the most complex initial magnetic field and consider non-ideal versions of the s5 and r5a simulations. 

In Fig. \ref{visc_heating}, we show the cumulative volume integrated viscous heating for a selection of simulations. Unsurprisingly, the viscous heating is larger in the s5 simulation with a Reynolds number of $10^3$ (solid red curve) than in the case with the Reynolds number $10^4$ (dashed red curve). By comparing the purple and red curves, we also see that for the same Reynolds number, there is significantly more heating for the s5 simulation than for the r5a simulation. Indeed, there is still more heating in the s5 simulation with a Reynolds number of $10^4$ than in the r5a simulation with a Reynolds number that is an order of magnitude larger. This is despite the wave drivers injecting a similar amount of Poynting flux (see Fig. \ref{Poynt_flux}). Previously (in Fig. \ref{rd_vort_plot}), we observed that the small scale formation rate is significantly reduced for the r5a simulation and this explains the lower viscous heating rate in this case. 

\begin{figure}[h]
  \centering
  \includegraphics[width=0.45\textwidth]{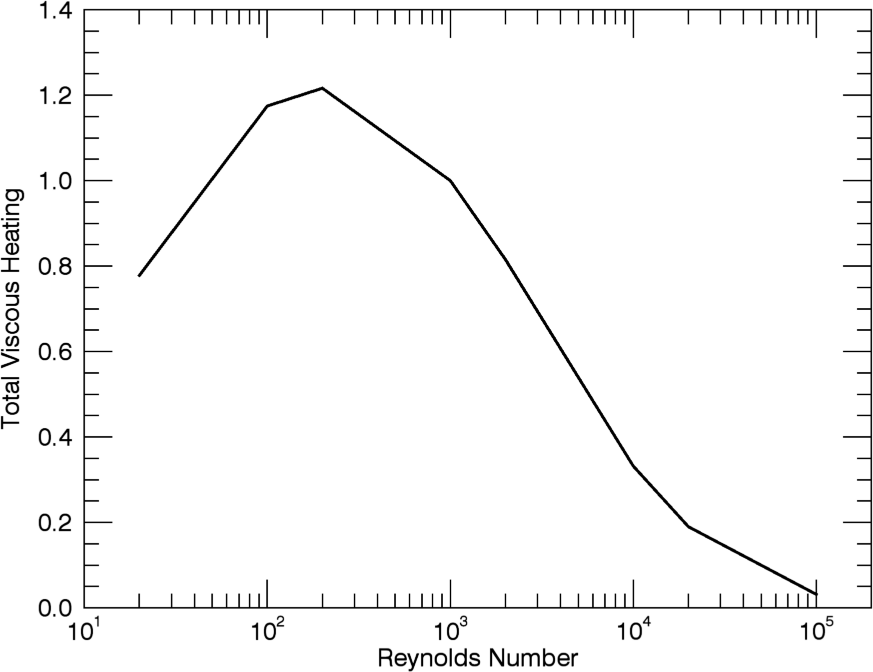}
  \caption{Total, time-integrated, viscous heating as a function of Reynolds number. Here we have normalised by the same value as for Fig. \ref{visc_heating}, the total viscous heating in the simulation with a Reynolds number of $10^3$.}
  \label{vh_Re}
\end{figure}

In Fig. \ref{vh_Re}, we show the total viscous heating that occurs for simulations with various Reynolds numbers. For comparison, the values shown correspond to the end times of the curves in Fig. \ref{visc_heating}. For the most part, we observe lower heating for higher Reynolds numbers, as expected. However, at low Reynolds numbers ($< 2\times 10^2$), the total heating increases with the Reynolds number. This is explained by considering the Poynting flux injected by the wave driver in these cases. In highly dissipative conditions, waves experience significant damping before they return to the driven boundary. As such, resonances are unable to form and the injected Poynting flux is reduced. We note that such high Reynolds numbers are unlikely to be relevant in the corona but such effects are important if wave energy is found to dissipate more rapidly than is currently expected.

\begin{figure}[h]
  \centering
  \includegraphics[width=0.45\textwidth]{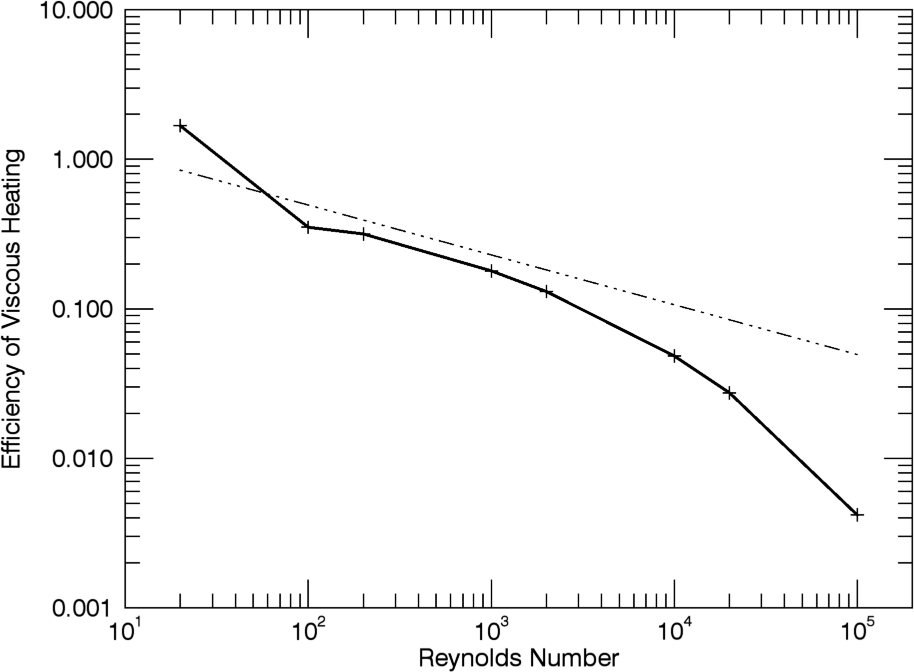}
  \caption{Rate of conversion from mechanical (kinetic and perturbed magnetic) energy to thermal energy  at $t = 2400 s$. The solid line shows the percentage of mechanical energy that is being dissipated every second for simulations with different Reynolds numbers. The dot-dashed line shows the gradient which should be obtained if the heating efficiency scales with $R_e^{-1/3}$.}
  \label{vh_efficiency}
\end{figure}

In Fig. \ref{vh_efficiency}, we isolate the heating efficiency for simulations with different Reynolds numbers. We display the rate of conversion from wave (kinetic and perturbed magnetic) energy to thermal energy as a percentage of the total wave energy present in the domain (solid line). The figure shows one instant in time ($t = 2400$ s) at the end of the simulation. Now we see that the efficiency of wave energy dissipation does indeed decrease with Reynolds number. We also show the dot-dashed line which has a gradient of $-1/3$ and demonstrates the relationship that would be expected if the heating rate was proportional to $R^{-1/3}$ as predicted by \citet{Heyvaerts1983}. We see that this approximates the gradient of the solid curve between Reynolds numbers of $10^2$ and $2 \times 10^3$. Since the relation is only valid for weak dissipation, the change of gradient for small Reynolds numbers ($< 10^2$) is not unexpected. The departure from the ${-1/3}$ power for large Reynolds numbers is likely due to numerical effects which become more significant for lower dissipation coefficients. In particular, the heating efficiency is reduced as more energy is lost numerically, and thus not converted into heat. We did not find evidence to suggest that for complex magnetic fields, the heating efficiency scales with the logarithm of the Reynolds number as suggested by \citet{Similon1989}, which would make phase mixing more viable as a potential coronal heating mechanism.

\section{Discussion and conclusions} \label{sec_Discussion}
In this article, we have presented the results of three-dimensional MHD simulations of waves propagating through a complex background magnetic field. We have shown that phase mixing and the non-linear interaction of counter-propagating waves can lead to a cascade of energy to small length scales and induce a significant increase in the rate of wave energy dissipation. These effects are sensitive to the complexity of the initial field, the magnitude of transport coefficients and the nature of the wave driver.

In a complex field, magnetic field lines have a wide range of natural frequencies due to variations in the Alfv\'en speed and in the length of field lines. Furthermore, different oscillation polarisations can have different natural frequencies depending on the geometry of the field. As such, for any wave driver with power in frequencies that are larger than the fundamental modes of field lines, we can expect to excite field line resonances, provided that the field is sufficiently complex. These resonances lead to the formation of large amplitude, non-linear waves which, in turn, excite fast-like perturbations within the domain. The existence of resonances increases the Poynting flux through the simulation boundary and, for the wave amplitudes considered within this study, could allow enough energy to be injected to balance losses in the Quiet Sun. 

However, for lower frequency wave drivers, resonances are unable to form and the Poynting flux is not sufficient to power any significant coronal heating. This remains a significant problem for wave heating models as observations suggest the majority of oscillatory power exists on time periods greater than the Alfv\'en travel time of many coronal loops \citep[e.g.][]{Morton2016}. Certainly, velocity amplitudes on the order generated in resonances within the above simulations are not currently observed. 

Within this article, we have focussed on the effects of phase mixing on the disruption of MHD wave fronts. However, since the system is 3-D, and the wave modes are non-linear, we can expect Alfv\'en wave turbulence \citep{Kraichnan1965, Howes2013} to also lead to the formation of small scales. Indeed, the volume integrated currents and vorticities discussed above inevitably includes contributions from a turbulent energy cascade. Alfv\'en wave-induced turbulence has been discussed in the context of coronal heating by \citet{VanBallegooijen2011, Verdini2012}. However, for the current parameter space, the increased small scale formation rate associated with the turbulent energy cascade remains insufficient to provide significant coronal heating. 

The development of MHD turbulence is sensitive to the perpendicular length scale and may be further encouraged by the introduction of a horizontally varying wave mode \citep[e.g.][]{Verdini2010, Shoda2018}. The effects of introducing such a wave were considered in the context of a single pulse by \citet{Howson2019}. However, in this case, the deformation of wave fronts due to phase mixing ensures that counter-propagating waves have significantly different horizontal wave numbers regardless of the spatial structure of the velocity driver. As such, a detailed consideration of the effects of the horizontal form of injected waves is unnecessary. Additionally, larger amplitude waves are able to induce greater MHD turbulence and plasma heating in the magnetic toplogy discussed here. However, the velocities obtained within the simulations discussed above are already larger ($\sim 150 \text{ km s}^{-1}$) than those currently observed in the solar corona. As a result, a consideration of higher amplitude driving cannot be physically motivated.

For the simple wave driver in these simulations, there are essentially three possible outcomes for the injected energy; (1) it is dissipated before returning to the driven boundary (following reflection), (2) it is stored within the corona in the form of large amplitude resonances (before being dissipated), and/or (3) it returns to the driven boundary and the driver is as likely to remove the energy as it is inject new energy (or else resonances must form). Since very large amplitude resonances have not been observed within the solar atmosphere, it seems reasonable to discount option (2). As option (3) implies wave energy is unimportant for heating the corona (which is certainly possible), only option (1) allows for waves to be a significant contributor to the coronal energy budget. However, for the generalised phase mixing considered within this article, we see that even with the inclusion of a viscous dissipation term that is many orders of magnitude larger than typically expected for the solar corona, significant heating is only produced when large resonances form (case s5 and r5a). Without these large resonances (case r5b), there is insufficient wave energy to allow appreciable heating to occur. Even when high amplitude field line resonances do form, for the field configurations considered here, the cascade of energy to dissipation scales is too slow to balance energy losses within the Quiet Sun.

Of course, it remains possible, that if we had considered a more complex field, with finer scales and larger gradients in the local Alfv\'en speed and field line length, greater wave energy may be extracted. However, in these genuinely 3-D regimes, wave energy is not confined to individual field lines and is able to propagate across field lines in the form of fast-like waves. Indeed, this is particularly relevant when the spatial gradients (e.g. in the total pressure) become large. This inhibits the rate at which energy can cascade to small scales and may explain why we did not obtain a heating efficiency that scales with $\log(R_e)$. Indeed, \citet{Parker1991} found only weak heating was possible for genuinely 3-D initial conditions (albeit in an open field coronal hole). Furthermore, as the field becomes more complex, we should expect significant energy to be released by the dissipation of currents in the background field and not necessarily through wave heating. Therefore, in sufficiently complex fields, it is likely that the contribution of waves to the energy dissipation budget would be small.

\vspace{1cm}

{\emph{Acknowledgements.}} The authors would like to thank Prof. Alan Hood, Prof. Peter Cargill and Dr Andrew Wright for their helpful ideas and discussions. The research leading to these results has received funding from the UK Science and Technology Facilities Council (consolidated grants ST/N000609/1 and ST/S000402/1) and the European Union Horizon 2020 research and innovation programme (grant agreement No. 647214). IDM acknowledges support from the Research Council of Norway through its Centres of Excellence scheme, project number 262622. JR acknowledges the support of the Carnegie Trust for the Universities of Scotland.

\bibliographystyle{aa}        
\bibliography{CF_WH.bib}           

\end{document}